\documentclass[
    ,final            
  ]
  {aipproc}

\layoutstyle{8x11double}
\newcommand{\beq}{\begin{equation}}
\newcommand{\eeq}{\end{equation}}
\begin{document}

\title{Origin of neutron star magnetic fields}


\author{H.C. Spruit}{
  address={Max-Planck-Institut f\"ur Astrophysik, Postfach 1317, 85741 Garching Germany}
}

\begin{abstract}
Possible origins of the magnetic fields of neutron stars include inheritance from the main sequence progenitor and dynamo action at some stage of evolution of progenitor. Inheritance is not sufficient to explain the fields of magnetars. Energetic considerations point to differential rotation in the final stages of core collapse process as the most likely source of field generation, at least for magnetars. A runaway phase of exponential growth is needed to achieve sufficient field amplification during relevant phase of core collapse; it can probably be provided by a some form of magnetorotational instability. Once formed in core collapse, the field is in danger of decaying again by magnetic instabilities. The evolution of a magnetic field in a newly formed neutron star is discussed, with emphasis on the existence of stable equilibrium configurations as end products of this evolution, and the role of magnetic helicity in their existence. 
\end{abstract}

\maketitle

\section{Sources of magnetic field}

At all stages of stellar evolution some percentage of stars  appears to be strongly magnetic: the magnetic A- and B -stars, magnetic white dwarfs, and neutron stars. Do these fields have a common origin or are different mechanisms at work? What kinds of mechanisms do we know of, and which would be the most plausible one for neutron stars? 
Three possible origins have been discussed for pulsar magnetic fields:
\begin{itemize}
\item{} the fossil field hypothesis, also called `flux conservation',
\item{} fields generated by  a dynamo process at some stage in the evolution of the progenitor star,
\item{} the thermomagnetic effect in the neutron star crust.
\end{itemize}
The last one will not be discussed here, since the effect is probably not enough to produce
pulsar and magnetar fields of the required strength. See, however, Ho et al. (2004).

\subsection{Fossil fields}
The simplest and most popular hypothesis, flux conservation, is that neutron star magnetic fields are simple remnants of the field of their main sequence (MS) progenitors. How does this work out quantitatively? The strongest fields known in MS stars are around $10^4$G. For a MS progenitor of radius $4\,10^{11}$cm, (10M$_\odot$) the star would contain a magnetic flux of $\sim 5\, 10^{27}$ G\,cm$^{2}$. A neutron star with the same flux would have surface field strength of $5\,10^{15}$G, sufficient for a magnetar. But this is an unrealistically optimistic estimate. The neutron star contains only some 15\% of the progenitor's mass, and the inner $1.4$M$_\odot$ of the progenitor occupy only some 2\% of the star's crossection. If the field is roughly uniform, the flux contained in the core would correspond to $10^{14}$G on a neutron star, insufficient for a magnetar. 

The second problem is statistics: only a very small fraction of the progenitors has magnetic fields as large as $10^4$G. Magnetars, on the other hand, are born frequently. As shown in P.\ Wood's talk (this volume), their birth rate is probably comparable to that of normal neutron stars. 

The fossil hypothesis is therefore not very attractive, at least not for magnetars. Magnetar fields might have a different origin than normal pulsars, of course, but that raises the obvious question why a process operating in magnetar progenitors could not work, at lower strength, in those of pulsars as well.

In any case, the `flux conservation' of a fossil field would have interesting consequences for the rotation of pulsars, since this scenario implies that a strong magnetic field is present in the star during its entire evolution. This includes the giant/supergiant phases in which the envelope rotates very slowly. Since this envelope also contains almost all the moment of inertia of the star, any 
coupling between core and envelope will quickly spin the core down to the same low rotation rate. The internal magnetic field needed to explain the magnetic flux of pulsars is more than enough to provide such coupling. Except in the very latest stages of evolution, when the Alfv\'en travel time through the star becomes longer than the evolution time scale, the core will corotate with the envelope. The angular momentum left in the core is then much too low to explain to rotation rates of pulsars. 

This would disqualify the fossil field hypothesis unless the rotation of pulsars is due to something else, for example `birth kicks', as proposed by Spruit and Phinney (1998). Either the angular momentum, or the magnetic field of pulsars can be fossil, but not both. 

Birth kicks would generically lead to correlation between spin axis and direction of proper motion (independent of whether the magnetic field is a fossil or not). Interestingly, such correlations are now being found (Rankin 2006). As in the case of the Crab and Vela pulsars, there appears to be a general preference for alignment of proper motion and spin axis on the plane of the sky. In the analysis of Spruit and Phinney, this means that the kicks are of relatively long duration, long compared with the spin period of the (proto-)neutron star at the time of the kicks (Romani 2005, Wang et al. 2006).

\subsection{Dynamos}
A magnetic field might be generated in the core of a SN progenitor at some earlier stage of its evolution, and subsequently amplified during core collapse (for example simply by `flux conservation'). Possible processes are:

\subsubsection{Convection}
Large scale magnetic fields generated by dynamo processes are traditionally associated with convective zones.  Field strengths of such dynamos are often estimated by assuming equipartition of magnetic energy density with the kinetic energy density of convection, $B^2=4\pi\rho v^2$ where $v$ is the convective velocity. The energy flux in convection is $F\approx\rho v^3$, and if this carries the star's luminosity $L$, the equipartition field strength is found as
\beq 
B_{\rm e}\approx M^{1/6}L^{1/3}r^{-7/6}\approx 10^9\,G\,({M\over M_\odot})^{1/6} L_{38}^{1/3} r_8^{7/6},\label{beq}
\eeq
where the mean density $3M/4\pi r^3$ has been assumed for $\rho$. Compressing such a field from $r=10^8$ to $10^6$ cm would yield $B=10^{13}$G. This would be sufficient for a normal pulsar.

It is not at all clear that this is a reasonable estimate, however. In the Sun, the {\em large scale} (dipole) field is some 300 times smaller than equipartition with convection would suggest. If this carries over to a stellar interior, convection-generated fields in the progenitor would be insufficient for producing pulsars fields.

It is also quite unclear if a field generated by a convective zone, even if it were strong enough, would leave a permanent dipole moment in the core when it becomes stably stratified again. Does the magnetic field generated in a convective region simply retreat, together with the convection, or can it leave a net magnetic field behind? Since expansion and retreat of a convective zone is slow compared with the characteristic dynamo reversal time, it is likely to leave behind, if anything a series of magnetic zones of different orientation instead of a net large scale field (much like the famous magnetic stripes along midocean ridges in the earth's crust).

\subsubsection{Field generation in stable zones}
Convection is not an essential ingredient in magnetic field generation in stars. The energy of differential rotation is sufficient. `Closing of the dynamo loop' can be achieved by instabilities in the magnetic field itself, in much the same way as `magnetorotational' fields  are generated in accretion disks (Hawley et al. 1995). 

Contrary to conventional wisdom, field generation can thus take place also in stably stratified zones of stars (Spruit 2002, Braithwaite and Spruit 2006). The stable stratification, however, strongly limits the radial length scale on which the process operates (just as it does with hydrodynamic processes, cf. Zahn 1983, 1992). For this reason it is unlikely to produce a large scale field. 

As in the case of convection it is unlikely that such a process would produce a net dipole moment of much significance for the neutron star descendant. It would, however, be of critical importance for the evolution of the angular momentum distribution in the progenitor, since even weak magnetic fields with small radial length scales can exert torques that are significant on evolutionary time scales (Heger et al. 2005).

\subsubsection{Neutrino-driven convection}
The convective velocities in the above estimates are driven by the (radiative) luminosity of the star. Much more powerful convection takes place during core collapse, driven by the much higher neutrino flux. The luminosity is then of order $10^{52}$ erg/s, the size of the core some $3\,10^6$, which yields  (from eq. \ref{beq}) $B_{\rm e}\sim 10^{15}$ G, i.e. in the magnetar range (Thompson \& Duncan 1993). However, if the actual dipole field generated is as small, compared with this equipartition number, as it is in the Sun, it would be 300 times smaller. Even after contraction from $3\,10^6$ to $10^6$cm, this would produce a field of only $10^{14}$ G, marginal for a magnetar, though sufficient for ordinary pulsars.

\subsection{Magnetic fields during collapse: energy estimates}

The arguments above suggest that field generation in the progenitor followed by compression during core collapse is not the most promising scenario for the production of pulsar fields. The field generation processes in the progenitor are important for providing a `seed field', but it still has to be amplified  by a large factor. 

Suppose a useful initial magnetic field has somehow been generated in the pre-SN core. If the collapse proceeds with flux conservation, $B\sim 1/R^2$, the magnetic energy $E_B\approx B^2R^3$ increases as $1/R$, i.e. as a constant fraction of the gravitational binding energy $E_{\rm G}$ of the core, $E_{\rm G}\sim GM^2/R$. Simple flux conservation therefore cannot increases a dynamically insignificant initial field into a dynamically significant field. A field of of $10^{15}$ G in the newly formed neutron star ($R=10^6$) would require a field of $10^{11}$G in a pre-collapse core of $R=10^8$ cm. As I have argued above, neither a fossil field nor an internally generated (pre-collapse) field is likely to be this high.

It is possible to do better, at least in simple energetic terms, by exploiting differential rotation of the star. Under angular momentum conservation the rotational energy varies as
\beq E_\Omega={1\over 2}I\Omega^2={1\over 2}J^2/I\sim 1/R^2.\eeq
This increases more rapidly than the gravitational energy. 
For the generation of a magnetic field,  only the energy fraction in {\em differential} rotation is relevant:
\beq 
E_{\Delta\Omega}=({\Delta\Omega\over\Omega})^2E_\Omega\sim 0.1E_\Omega,\sim{1/R^2}\label{eom}
\eeq
where I have assumed $\Delta\Omega\approx 0.3\Omega$ as a plausible rate of differential rotation. 

Assuming there is a process that can convert all energy of differential rotation into magnetic energy $E_B=B^2R_{\rm NS}^3\sim E_{\Delta\Omega}$, this would produce a maximum field strength inside the star
\beq B_{\rm max}\approx 10^{17}{\Delta\Omega\over\Omega}{\Omega_{\rm NS}\over\Omega_{\rm K}},\label{bmax}\eeq
where $\Omega_{\rm NS}$ is the spin of the neutron star end product and $\Omega_{\rm K}$ the Kepler frequency at the neutron star surface. The observable surface field strength would be lower than this internal field, and the conversion of differential rotation into magnetic energy less than 100\% efficient. These efficiency factors are hard to guess, but it suggests that field strengths of the order of a magnetar field are plausible, provided the collapsing core contains enough angular momentum. 

How large can the angular momentum of the collapsing core be? An upper limit is set by the average energy of observed core collapse supernovae, of the order $10^{51}$ erg. If the rotational energy of the neutron star formed is larger than this, and the the star strongly magnetic, spindown by pulsar emission will be so fast that the rotational energy is dumped already into the supernova itself. This sets an upper limit of $10^{51}$ erg to the rotational energy, corresponding to a rotation period of 4ms for a neutron star of 1.4M$_\odot$, or $\Omega_{\rm NS}/\Omega_{\rm K}\approx 0.1$. The maximum field strength from (\ref{bmax}) is then $3\,10^{15}$ G, assuming again $\Delta\Omega/\Omega=0.3$. This shows that magnetar  field strengths, up to $10^{15}$, are possible, but require a fairly efficient conversion of differential rotation into magnetic energy. 

A way of estimating the angular momentum available during core collapse is to follow the evolution of the internal rotation in the progenitor up to core collapse. This requires faith in one's quantitative understanding of the angular momentum transport  processes in the star. Internal torques due to magnetic fields are probably more important than purely hydrodynamic processes alone. Heger et al. (2005) have computed the initial spins of neutron stars with the prescription for magnetic torques from Spruit (2002). Rotation periods in the range 8-10ms were obtained, somewhat independent of the initial (main sequence) spin. This is comfortably below the upper limit given above, but still rapid enough for magnetar fields of $10^{15}$G. 

\subsubsection{Consequences and extrapolations}
If differential rotation is an essential ingredient for the production of a strong magnetic field, as suggested above, at least magnetars must be formed spinning rapidly. If a magnetar with $B=3\,10^{14}$G is formed in this way, and becomes visible through the supernova debris after a couple of years, say, it would spin at about $P\sim 0.1$s, and would be more luminous than the Crab pulsar. It is not clear if this is compatible with observations of supernovae in nearby galaxies.

On the other hand, it is clear from pulsars with independent age estimates (independent from the spindown time scale) that normal pulsars are formed with a range of initial spins, from $\sim 15$ms (Crab) to 400 ms (Camilo et al. 2007, Gotthelf \& Halpern 2007, see also Gotthelf, this volume). This shows that there is either a significant spread in angular momentum of the collapsing core, or that an efficient post-collapse spindown process operates, such as friction against a fall-back disk (cf. Li \& Jiang 2007, Wang et al. 2007). 

Some of the slow-born pulsars have rather weak fields (Camilo et al. 2007, Gotthelf \& Halpern 2007). This  would agree with angular momentum of the pre-SN core playing an important role in determining the field strength of a neutron star, as suggested above.

The rapid initial spin required for magnetar fields can be extrapolated to higher rotation rates. At sufficiently rapid rotation, the spindown of the nascent, highly magnetized neutron star could power a Gamma-ray burst. This magnetar-GRB scenario has recently become popular, cf. Kommissarov \& Barkov (2007), Bucciantini et al. (2007), Yu \& Dai (2007) and references therein. 

\section{Requirements on the field-production process}
\label{ingred}
\begin{figure}[h]
\hfil{\includegraphics[width=0.7\hsize]{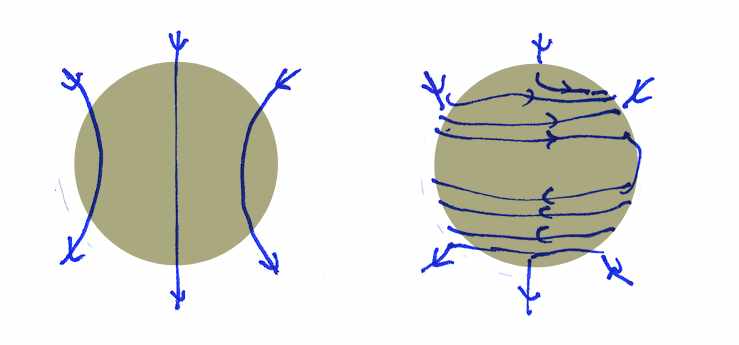}}\hfil
\caption{\label{wind} `Winding up' of a poloidal field by differential rotation. The field lines are stretched into a stronger azimuthal field inside the star. The surface field remains unchanged.}
\end{figure}

To explain the strength of the dipole component of a pulsar magnetic field, not just any field generated inside in a SN core will do. An illustrative example is the `winding-up' of an initial magnetic field (or `seed field') by differential rotation. 

Suppose the field is initially a purely poloidal field ${\bf B}_{\rm p}$ (field lines in meridional planes, no azimuthal field component), as sketched in fig \ref{wind}. Differential rotation stretches the field lines in azimuthal direction, creating an azimuthal field component $B_\phi$ that increases linearly with time, $B_\phi\sim B_{\rm p}\Delta\Omega t$. The surface field, and its dipole component remain {\em unchanged} in this process. 

\subsection{Stratification}

\begin{figure}
\hfil{\includegraphics[width=0.9\hsize]{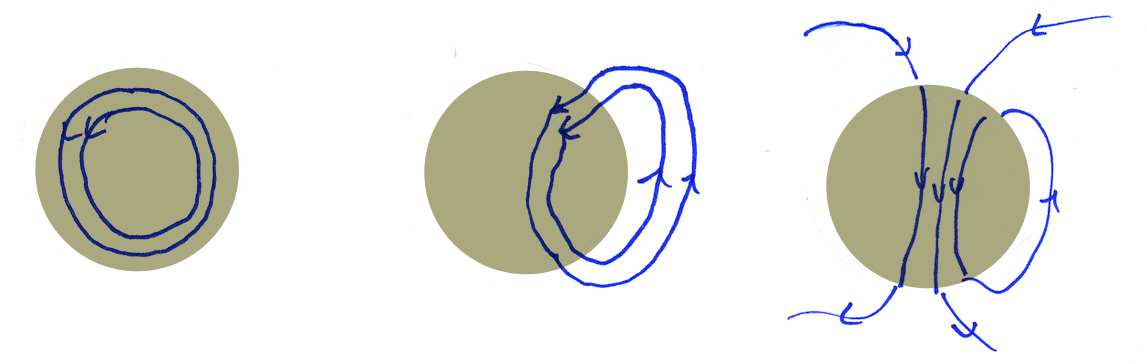}}\hfil
\caption{\label{rise} Magnetic buoyancy of an azimuthal field loop causes part of it to rise, provided the field is strong enough to overcome the stratification. This creates a dipole component at the surface.}
\end{figure}

To produce an interesting pulsar dipole, something else has to happen: the strong internal field  $B_{\rm p}$ somehow has to be brought to the surface. The obvious process is magnetic buoyancy: a strong field reduces the gas pressure and density in it, so that loops of the azimuthal field tend to float upward against the direction of gravity (see fig \ref{rise}). The stretch of the loop that finds itself outside the star forms a vacuum field with a dipole moment. In this way an initially weak poloidal field can be amplified by differential rotation into a field with a strong dipole. 

In a stable stratification, a magnetic field can float to the surface only if it is strong enough to overcome the stratification. This is the case, approximately, when the Alfv\'en speed $V_{\rm A}$ exceeds $HN$, where $H$ the pressure scale height and $N$ the buoyancy frequency of the stratification. Once the neutron star has formed, the stratification of the neutron to proton density ratio is able to prevent fields of up to $10^{17}$ G from emerging by such a direct buoyancy instability. 

A magnetic field of $10^{15}$ G produced internally by differential rotation must therefore have reached the surface in an early stage of the formation of the neutron star, when the buoyancy frequency was a percent or less of the final value. Buoyant rise is aided, however, by the presence of a dense neutrino field (see Thompson \& Murray 2001 for a detailed discussion).

\subsection{Instability of poloidal fields}

\begin{figure}[h]
\begin{minipage}[c]{1\hsize}
\centering\includegraphics[width=0.6\hsize]{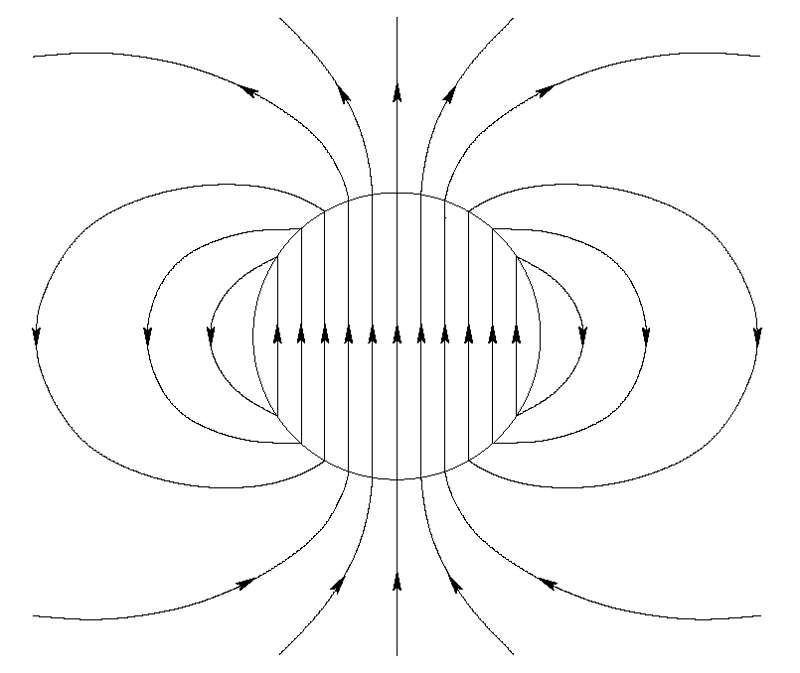}
\end{minipage}%
\end{figure}
\begin{figure}[h]
\begin{minipage}[c]{1\hsize}
\centering\includegraphics[width=1\hsize]{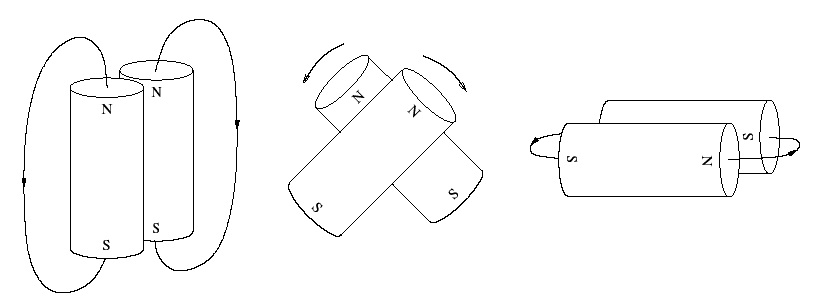}
\end{minipage}
\caption{\label{FRfig} Instability of a poloidal field in a star. (From Flowers and Ruderman, 1977)}
\end{figure}

A poloidal field, such as that created by the buoyant process of fig. \ref{rise}, cannot be the dipole field seen in pulsars, since it is highly unstable. All purely poloidal fields in a star are unstable (Wright, 1973, Markey and Tayler 1974, Flowers \& Ruderman 1977). A simple example is that of Flowers and Ruderman, shown in fig \ref{FRfig}. In this example a uniform field inside the star connects to a vacuum field outside. This external field is that of a point dipole centered on the star (cf.\ Jackson E\&M). Cutting the star in two halves without cutting field lines is possible along the axis of the field. Rotating one half in place by 180$^\circ$ does not change any of the energy components in the interior of the star (thermal, gravitational, magnetic). But the external magnetic energy changes, in the same way the energy of two parallel bar magnets changes when one is rotated by 180$^\circ$. The instability proceeds on the time it takes an Alfv\'en wave to cross the star.

\begin{figure}[h]
\hfil\includegraphics[width=0.65\hsize]{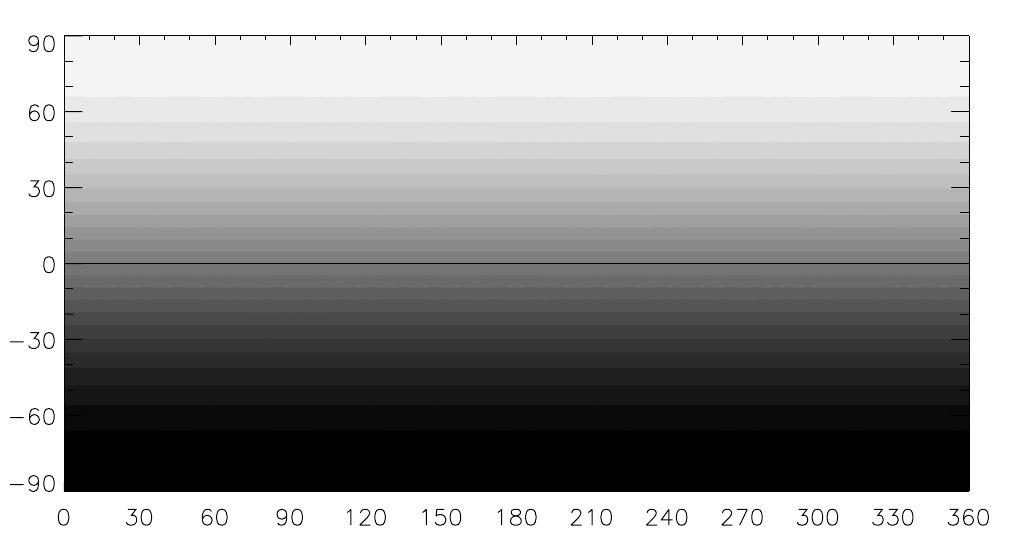}\hfil\includegraphics[width=0.35\hsize]{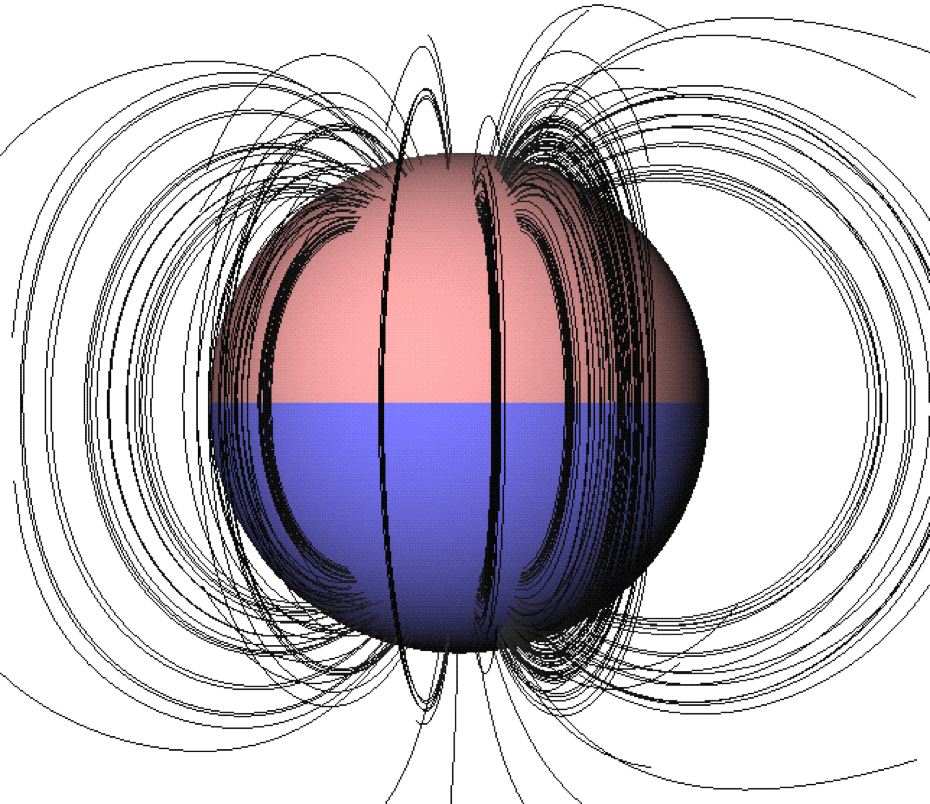}\hfil
\end{figure}
\begin{figure}[h]
\hfil\includegraphics[width=0.65\hsize]{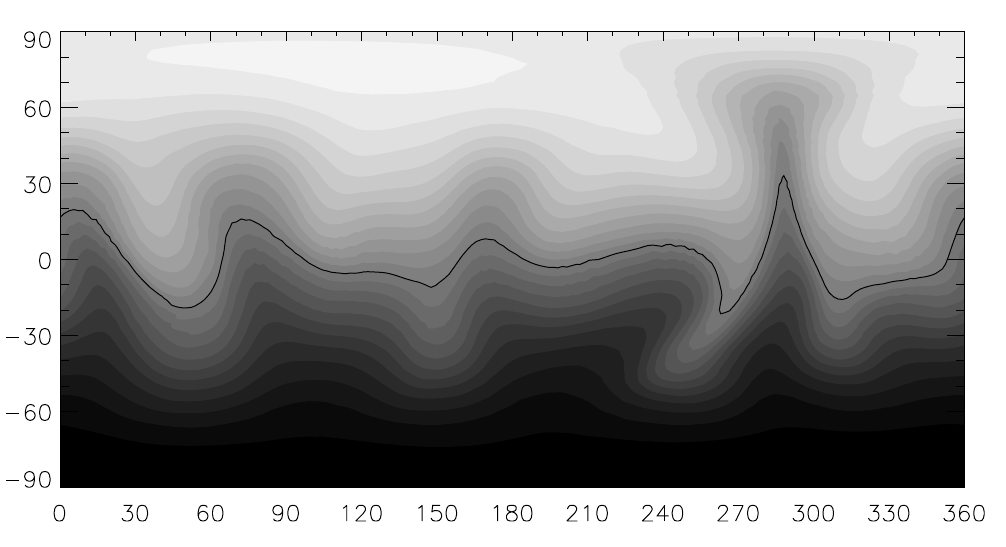}\hfil\includegraphics[width=0.35\hsize]{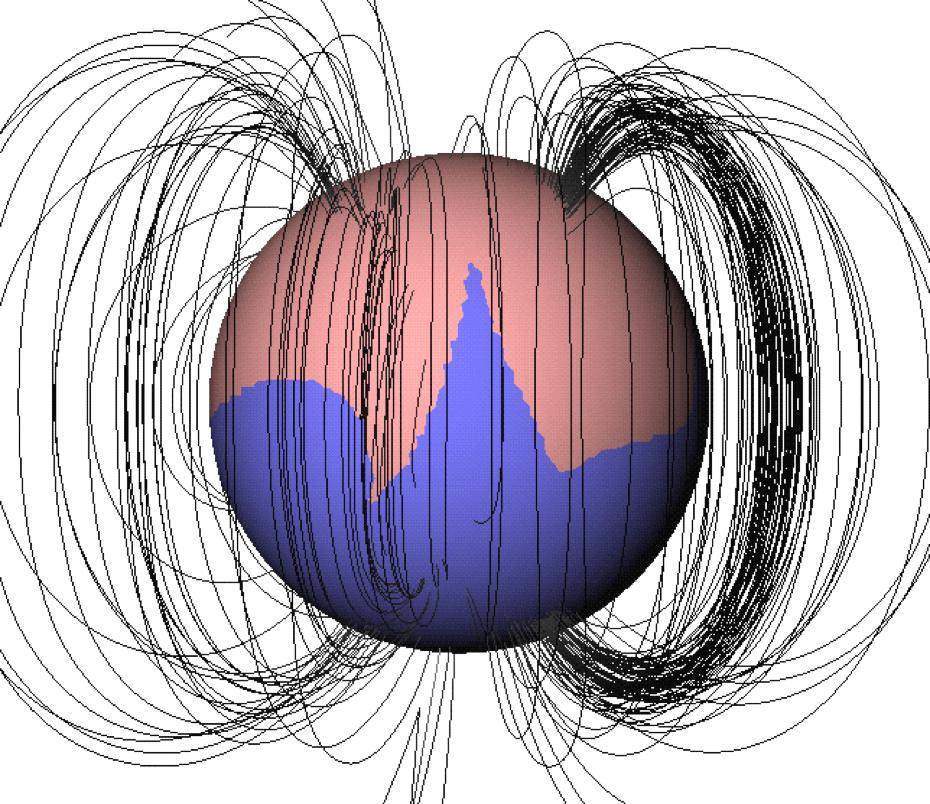}\hfil
\end{figure}
\begin{figure}[h]
\hfil\includegraphics[width=0.65\hsize]{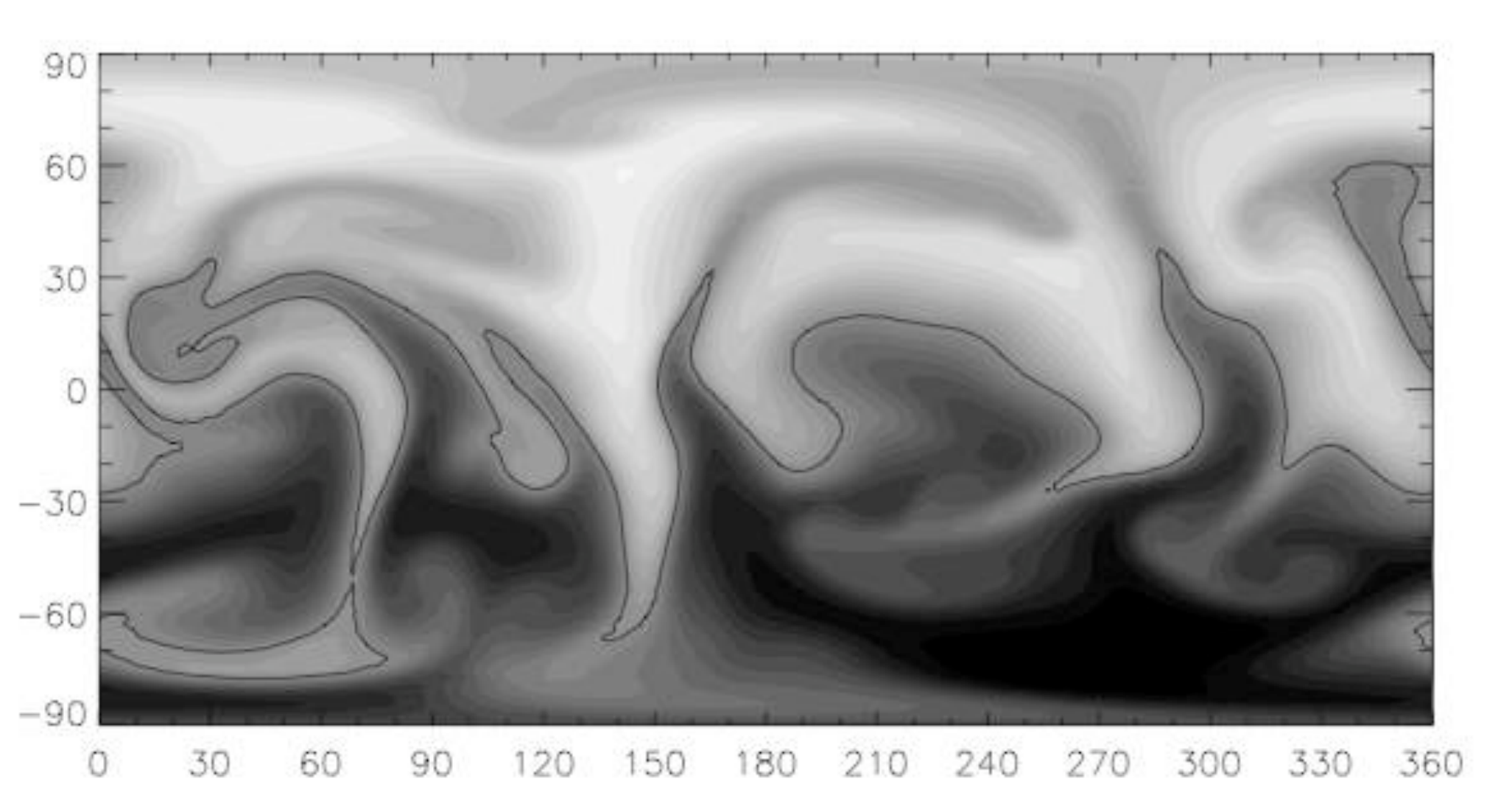}\hfil\includegraphics[width=0.35\hsize]{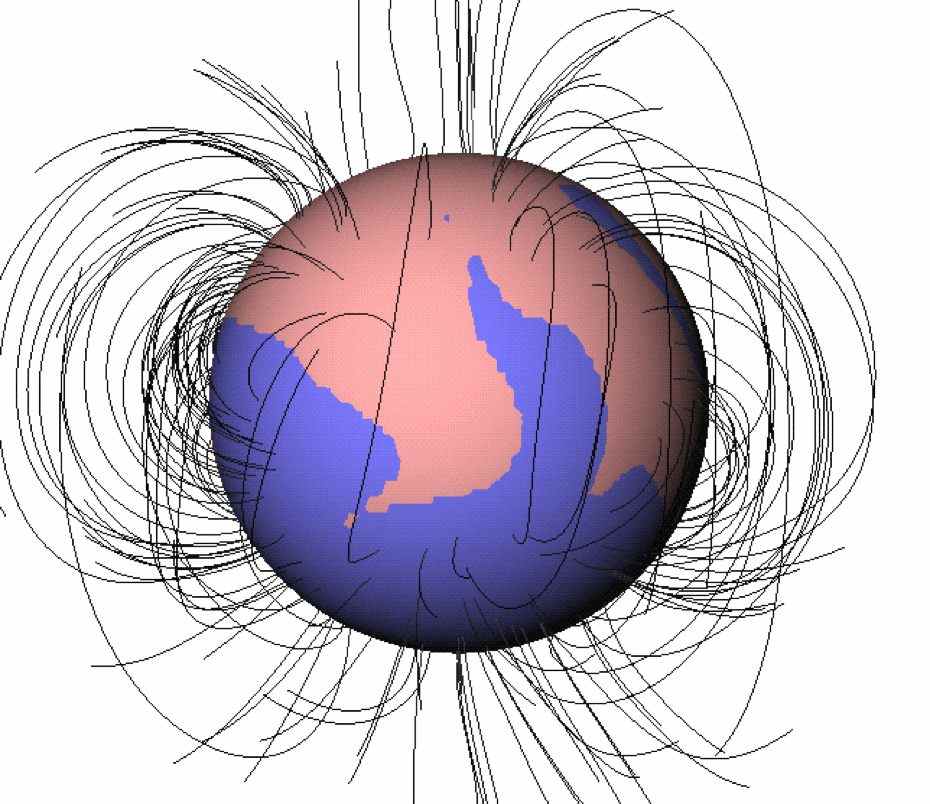}\hfil
\caption{\label{braith} Numerical simulation of Flowers-Ruderman instability in a star.  Left panels: radial component of the field strength on the star's surface, in the initial dipolar state and at two times during the nonlinear evolution of the instability. The magnetic equator (radial component zero) is shown as a dark line (from Braithwaite and Spruit 2006)}
\end{figure}

How would this bar magnet process work in practice, in a gravitating fluid sphere like a star? This can be answered in some detail with numerical simulations. Fig. \ref{braith} shows the result of calculations by Jonathan Braithwaite. The star does not flip hemispheres as fig. \ref{FRfig} suggests. Instead, the instability sets in at a smaller azimuthal length scale. The evolution of the surface field distribution resembles the mushroom-like mixing patterns observed in Rayleigh-Taylor instability. Both properties are characteristic of interchange type instabilities. [The radial field component at the surface shows the instability nicely, but of course the entire interior of the star takes part in it.] 

In these simulations, the field decayed completely, within the numerical uncertainties. This can be understood from the properties of {\em magnetic helicity}, discussed below. The conclusion is that buoyant rise of a loop of strong field is not enough, at least not in the simple form of fig \ref{rise}. At least one more ingredient is needed to end up with a magnetic field configuration that is useful for explaining magnetar fields.

\subsection{The role of magnetic helicity}
\begin{figure}[t]
\hfil{\includegraphics[width=0.8\hsize]{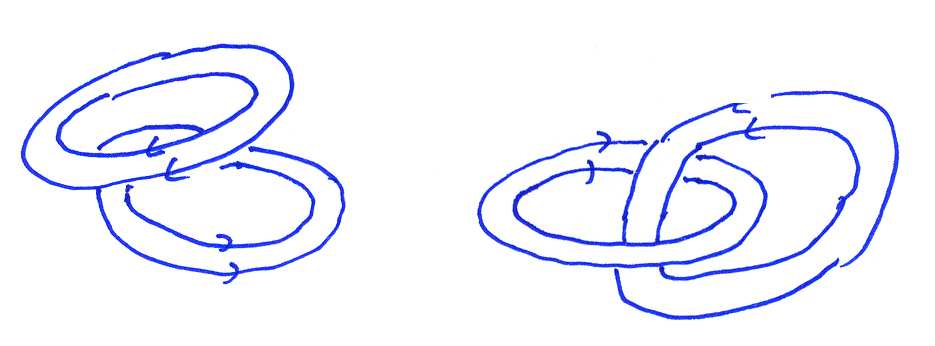}}\hfil
\caption{\label{helici} Magnetic helicity. A configuration consisting of separated loops of untwisted magnetic field (left sketch) has zero helicity. A `linked poloidal toroidal' field made of the same loops (right) is helical.}
\end{figure}

The magnetic helicity $\cal H$ of a field configuration is a global quantity,
\beq {\cal H}=\int {\bf B\cdot A}~{\rm d} V,\eeq
where $\bf A$ is a vector potential of $\bf B$, and the integral is over the volume of the magnetic field. [The gauge dependence of $\bf A$ is physically significant. It causes helicity to be global: $\bf B\cdot A$ has no meaning as a local `helicity density'].  In a perfectly conducting fluid with fixed boundary conditions magnetic helicity is a conserved quantity. In practice perfect conduction is not a realistic situation, since rapid reconnection can take place even at very high conductivity, especially in a dynamically evolving field configuration. Nevertheless, in laboratory experiments helicity is often observed to be at least approximately conserved. 

\begin{figure}[h]
\hfil{\includegraphics[width=0.6\hsize]{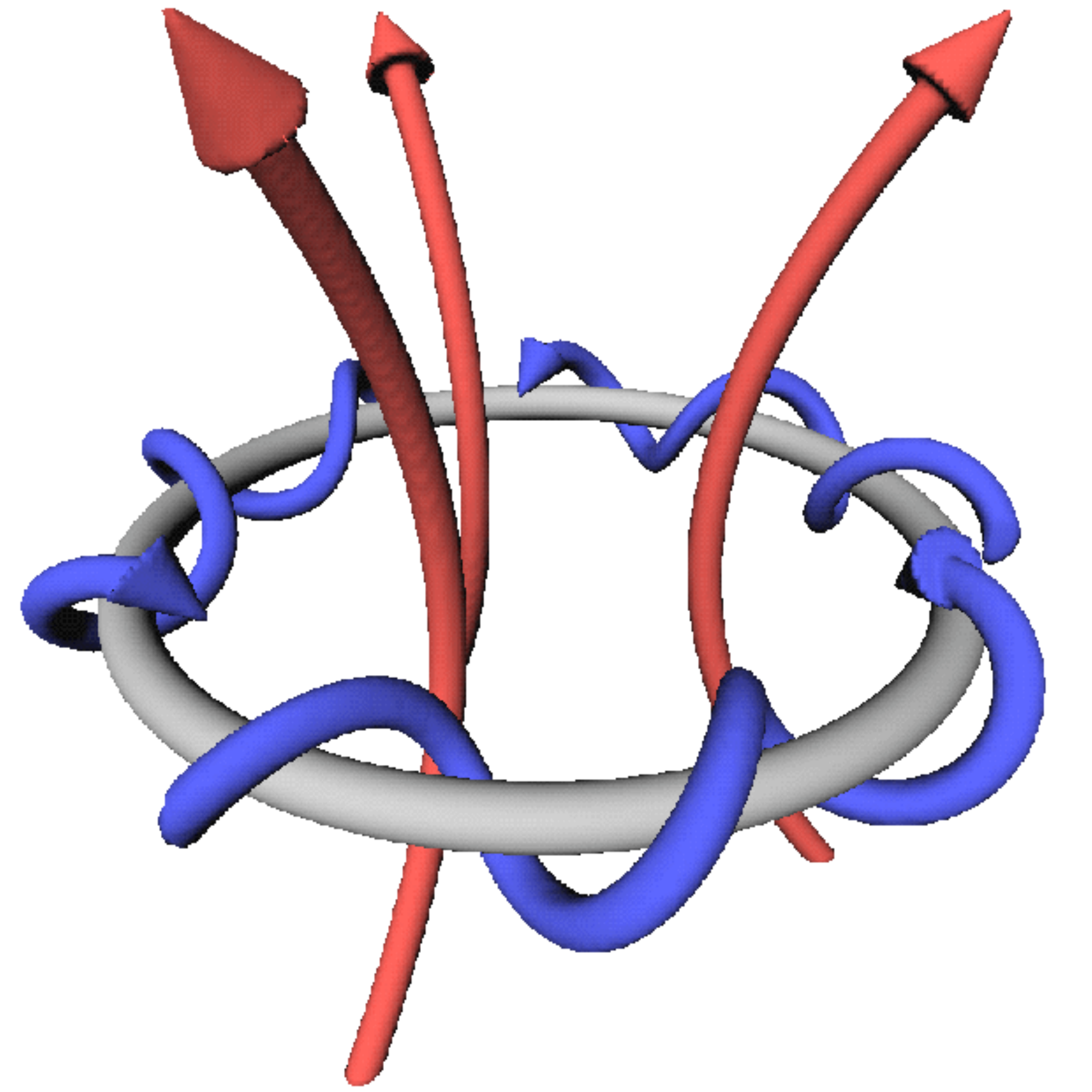}}\hfil
\caption{\label{donut} Equilibrium found as end state of the decay of a random field with finite helicity in a stably stratified star (idealized sketch). The surface field is determined by the  poloidal field lines feeding through the hole of the torus. On their own they would be unstable as in fig \ref{FRfig}; they are stabilized by the twisted torus surrounding them. (From Braithwaite and Nordlund 2006)}
\end{figure}

The consequence is the existence of stable equilibrium configurations. To the extent that helicity is conserved, a non-equilibrium or unstable magnetic field with a finite helicity cannot decay completely, since the helicity of a vanishing field is zero. An experiment like that of fig \ref{braith},  a field starting  with a finite amount of helicity therefore decays initially, but eventually settles into a stable equilibrium. [In laboratory plasma physics this process of settling is called Taylor relaxation]. 

Two factors are thus necessary for a stable magnetic equilibrium in a fluid star: a stable stratification (otherwise the field would just escape through to the surface by its buoyancy) and a finite magnetic helicity.

Stellar models with such equilibria have been found by Braithwaite and Nordlund (2006). They usually consist of an approximately axisymmetric torus of twisted field lines inside the star, and a bundle of field lines feeding through the `donut hole' and through the surface of the star, see fig. \ref{donut}. The existence of stable configurations of this general form in stars has been surmised several times in the past (Prendergast 1956, Kamchatnov 1982, Mestel 1984). It is quite possible that there also exist stable fields with more complex configurations.

\subsection{Field amplification in core collapse}

To make use of the energy of differential rotation (eq. \ref{eom}), a process has to be found that can convert it to magnetic energy, and can do so fast enough. Most of the rotational energy becomes available only towards the end of the collapse, when the stratification begins to develop its stable buoyancy and starts to suppress emergence of the field to the surface. This leaves a window of perhaps a few seconds for an observable surface magnetic field to grow from rotational energy. 

To see what this implies, approximate the collapse as a two stage process: an instantaneous collapse to neutron star size $R=10^6$ cm with angular momentum conservation and flux freezing, followed by a phase of field amplification by differential rotation. Consider first the simple winding-up process of fig \ref{wind}. With a nominal rotation period of 10ms, $\Delta\Omega/\Omega=0.3$, and a time available of 3s, the field can be amplified by a factor 100. To reach a field strength of $10^{15}$G, the initial field would have to be $10^{13}$, corresponding to $10^9$ in a pre-collapse core of $10^8$cm. This is marginally achievable with differential rotation-generated fields (Heger et al. 2005), but does not leave much room for inefficiencies. In addition, as shown in fig \ref{wind}, something more has to happen to convert a wound-up internal field into something observable (i.e. a field with a dipole moment at the surface). 

This something is likely to be an instability affecting the magnetic field. If the instability operates on a magnetic time scale $R/v_{\rm A} $, it will have noticeable effects only when this time scale becomes significantly shorter than our time window of a few seconds. If such an instability increases the poloidal field strength, the azimuthal field strength produced by winding-up in the continuing differential rotation will then increases more rapidly than before, and the instability will grow faster. The net effect is a runaway amplification. 

There are three instabilities that can in principle be involved in such a scenario. Buoyant instability (fig \ref{rise}) will be most effective when the stratification is convective; a Tayler instability (displacements nearly along equipotential surfaces, cf. Spruit 1999) is also effective in a stable stratification, and finally there is magneto-rotational instability (MRI) which requires only an outward decrease of the rotation rate. 

For the present discussion the main distinction is between buoyant and Tayler instability on the one hand, and MRI on the other. The first two grow on a magnetic time scale  $R/v_{\rm A} $, hence become effective only when the field has already been amplified to some extent by linear winding-up. MRI, understood as a form of (`magnetically enabled') shear instability, grows on a differential rotation time scale $1/\Delta\Omega$, hence would be expected to set in immediately. This gives MRI modes a clear advantage when time is of the essence. 

\subsubsection{Numerical simulations}
An example of such a process is given by the simulations of Akiyama et al. (2003), Ardeljan et al. (2005), Moiseenko et al. (2006). For an instability with a fixed growth rate (independent of the field strength) the time $t$ needed to reach a given amplitude $B$ from an initial field strength $B_0$ scales as $t\sim\ln(B/B_0)$. This is roughly consistent with the evidence given by Moiseenko et al. (2006) and suggests that some form of MRI is responsible. 

The standard case presented by Moiseenko et al. (a ratio of magnetic to gravitational energy of $10^{-6}$), corresponds to a wildly optimistic initial field strength ($\sim 10^{15}$G at $R=10^6$, or $10^{11}$ at $R=10^8$cm). But because of the logarithmic dependence on $B_0$ the process still works with a more realistic initial field $10^3$ times smaller. It is possible that it will work even better in 3 dimensions, because of the additional modes of instability available in 3D.

\subsection{The first three minutes}

Suppose a field of pulsar or magnetar strength has been formed during core collapse, and that the stratification of the proto-neutron star has become stable enough to prevent the field from escaping by simple buoyant rise. After this dynamical phase the field is still out of equilibrium or at least far from stable. At the same time the star is still fluid: a crust forms only after some 100s, and initially is probably more ductile and deformable than a cooled-down neutron star crust. The evolution of the field over the first few minutes is therefore critical for its survival as a pulsar or magnetar field.

The field will decay by whatever magnetic instabilities are available to it. Without rotation, these instabilities will operate on time scales of the order of the Alfv\'en crossing time through the star, $t_{\rm A}\approx 3{\rm s}~B_{14}^{-1}$. Such instabilities continue to exist in a rapidly rotating star because the Coriolis force is perpendicular to the velocity, hence does not contribute an additional energy barrier to be overcome by instability (like a stable stratification does). 

The Coriolis force restricts the range of modes that are unstable when the rotation rate is larger than the Alfv\'en frequency $1/t_{\rm A}$. But usually there are some modes that remain unaffected by the Coriolis force (a trivial example being modes with displacements nearly parallel to the rotation axis). Most modes suffer only a reduction of their growth rate, characteristically by a factor $\Omega t_{\rm A}$, while some can actually become stable (Pitts and Tayler 1986). 

A simple survival estimate can be made by assuming that the growth of the relevant unstable modes is
reduced by the characteristic factor $\Omega t_{\rm A}$ (Flowers and Ruderman 1977). Without the helicity constraint, the field would then decay with time $t$ such that the growth time of the instability is of the order $t$: $t_{\rm A}^2\Omega\approx t$. After 200s, this yields as maximum surviveable field strength in the absence of a helicity constraint:
\beq B(200)\approx 5\,10^{12}\,{\rm G}~ P^{-1/2},\eeq
where $P$ is the spin period in s. For an initial spin of 10ms, fields of the order of observed pulsar strength could thus survive until the crust forms and saves them from further decay. 

Fields higher than $\sim 5\, 10^{13}$, i.e. magnetar fields, can survive only if their helicity is sufficient to reach a stable equilibrium. In fact, this holds independent of crust formation, since $10^{13}$G is about the maximum field strength believed to be supportable by a crustal lattice. For the survivability of magnetar-strength fields, the neutron star can be treated as fluid, with or without crust. [But the  crust  of course affects the actual decay of magnetar fields, modulating the energy releases into a series of starquakes.]

The effect of rotation on field decay can again be tested by numerical simulation. Braithwaite (2007) finds that rotation does indeed slow down the decay, but not quite in the simple way assumed above. Starting with the uniform internal field of the Flowers-Ruderman argument, oriented at some angle to the rotation axis, there is first a stage of decay which proceeds at the Alfv\'en crossing time, independent of the rotation rate. The implication is that this phase is due to the modes that are not affected by the Coriolis force. The further decay is slowed down by rotation, but within numerical uncertainty no stable equilibria were found, consistent with the vanishing helicity of the initial state. 

\section{Summary}

Three popular scenarios for the origin of  magnetic fields in pulsars and magnetars are i) the fossil field hypothesis, ii) fields generated internally in the progenitor, and iii) field generation during core collapse. Fossil fields are ruled out for magnetars (required field strength of the MS progenitor too large, number of magnetic MS progenitors too small). If a fossil field plays a role for normal pulsars, magnetic coupling between core and envelope of the progenitor will  remove the angular momentum from the core, producing neutron stars with extremely long rotation periods. The spin of pulsars must then be due to the `birth kicks' that also give rise to their proper motions. 

Internally generated fields, followed by core collapse under flux conservation, is unattractive because of the rather low field strengths that can plausibly be produced. The most plausible source of energy for field generation is the differential rotation during core collapse; it increases more rapidly than even the gravitational energy. This predicts that the most magnetic stars are also formed rotating intrinsically rapidly (but suffering rapid spindown by a pulsar wind already during the supernova, cf. Metzger et al. 2007). 

The most critical factor for the production of  a strong field is the finite time available in the final collapse phase, on the order of seconds. At earlier times in the collapse the energy in differential rotation is not yet high enough, at later times (when the neutrinos have left) there may still be enough energy in differential rotation, but the stratification has then become so stable that even fields of magnetar strength cannot reach the surface any more.

Combined with the lowish initial field strengths in the pre-collapse core, this implies the existence of a very efficient field amplification process during collapse. Linear winding-up in differential rotation is too slow except possibly for normal pulsars. Runaway growth of a magnetic field is possible by the combination of differential rotation with magnetically driven or shear-driven instabilities. Magnetorotational instability works fastest because it is essentially a shear-driven instability operating on the short time scale of differential rotation. The simulations of Akiyama et al. (2003) and Ardeljan et al. (2005) illustrate the feasibility of such a process.

Most simulations of magnetic core collapse (e.g. Shibata et al. 2006, Obergaulinger et al. 2006, Sawai et al. 2007, Burrows et al. 2007, see also the more analytical considerations in Uzdensky and MacFadyen 2007) start with rather optimistic assumptions on the strength and configuration of the pre-collapse field (an aligned dipole of $10^{11}$G at $R=10^8$ cm for example). Such initial states produce interesting results in themselves. But to explain magnetar fields, or  magnetically powered supernovae, let alone the magnetic field strengths needed to produce magnetically powered GRB jets, much more has to happen  than the flux freezing collapse of a dipole field.  It is likely that pre-collapse fields with more realistic configurations and strengths, and processes like those seen in Moiseenko et al. (2006), will produce magnetic fields that behave rather differently. In particular it is still an open question how the ordered, rotation-aligned magnetic field configurations are formed that work so well in producing jets (or if they form at all).

Since the field of the neutron star is formed in a highly dynamic event, it is likely to be far from equilibrium, or at least far from stable. For the same reason it is also likely to be dominated by higher multipoles rather than the dipole component that determines pulsar phenomenology.  For field strengths of $10^{13}$ G and less magnetic instabilities are slow enough for the field to survive until it can be anchored by the newly formed crust (rapid rotation helps here, since the Coriolis force reduces the growth rate of instabilities). 

Fields of magnetar strength decay rapidly, and are too strong to be anchored by the crust. Their apparent survival implies that they have found a stable equilibrium configuration within seconds after their formation. The formation of such stable equilibria from complex, strongly unstable initial fields can be demonstrated with numerical simulations. The key ingredients for the existence of such equilibria are i) the stable stratification of a neutron star, and ii) a nonvanishing magnetic helicity.

\begin{theacknowledgments}
I thank Chris Thompson, Mal Ruderman and Joseph Taylor for discussions during the conference, from which the present text has benefited. 
\end{theacknowledgments}

\end{document}